\documentclass[12pt]{iopart}

\usepackage{cite}

\usepackage{graphicx}

\usepackage{dcolumn}

\begin{document}

\title[Charged particle in a quadrupole field]{Bound states of a charged particle in the field of an electric quadrupole
in two dimensions}
\author{Francisco M. Fern\'andez}

\address{INIFTA (UNLP, CCT La Plata-CONICET), Divisi\'on Qu\'imica Te\'orica,
Blvd. 113 S/N,  Sucursal 4, Casilla de Correo 16, 1900 La Plata,
Argentina}\ead{fernande@quimica.unlp.edu.ar}

\maketitle

\begin{abstract}
The Schr\"odinger equation for a charged particle in the field of
a nonrelativistic electric quadrupole in two dimensions is known
to be separable in spherical coordinates. We investigate the
occurrence of bound states of negative energy and find that the
particle can be bound by a quadrupole of any magnitude. This
result is remarkably different from the one for a charged particle
in the field of a nonrelativistic electric dipole in three
dimensions where a minimum value of the dipole strength is
necessary for capture. Present results differ from those obtained
earlier by other author.
\end{abstract}

\section{Introduction}

Some time ago Alhaidari\cite{A07} discussed the problem of a
charged particle in the field of a nonrelativistic electric
quadrupole and calculated the minimum quadrupole strength that
allows the particle to be bound by the charge distribution. The
author chose the charge distribution of four fixed point charges
with zero total charge and dipole moment. The first nonzero
contribution to the multipole expansion is, therefore, the
quadruple term. Since this term is inversely proportional to the
square of the distance between the fifth charge and the origin of
the distribution the Schr\"{o}dinger equation is trivially
separable in spherical coordinates.

In some ways that problem resembles the most widely studied one of a charged
particle in the field of a nonrelativistic dipole moment in three dimensions%
\cite{CEFGC01} (and references therein). This simple model proved useful in
nuclear as well as in molecular physics\cite{CEFGC01} (and references
therein). In the latter field it has been used to predict the capture of an
electron by a polar molecule to produce an anion. I has been known from long
ago that the polar molecule cannot bind an electron unless the dipole is
greater than some critical value\cite{CEFGC01} (and references therein).
Camblong et al\cite{CEFGC01} gave a simple and smart proof, based on
dimensional scaling, of why the simple model is successful.

Alhaidari\cite{A07} arrived to a similar conclusion in the case of
the particle in the field of an electric quadrupole in two
dimensions. On revising his arguments we found that his
conclusions may not be correct. Since the problem may be of
physical interest we put forward our results in this paper. In
section~\ref{sec:bound_states} we solve the Schr\"{o}dinger
equation and derive the conditions for the capture of the charged
particle by the electric quadrupole. We split the Schr\"{o}dinger
equation into the angular and radial parts in the usual way and
show that the radial equation is different from the one derived by
Alhaidari. We also show that the standard solution of the angular
part yields results that are different from those obtained by that
author. As a result we obtain conditions for the capture that
considerably differ from those given earlier. In section~\ref
{sec:conclusions} we summarize the main results and draw
conclusions.

\section{Bound states}

\label{sec:bound_states}

The dimensionless Schr\"{o}dinger equation for a particle moving in a
potential $V(\mathbf{r})$ is
\begin{equation}
\left[ -\frac{1}{2}\nabla ^{2}+V(\mathbf{r})\right] \psi (\mathbf{r})=E\psi (%
\mathbf{r}).
\end{equation}
In two dimensions this equation is separable in spherical coordinates $%
0<r<\infty $ and $0\leq \theta <2\pi $ when
\begin{equation}
V(\mathbf{r})=V_{r}(r)+\frac{V_{\theta }(\theta )}{r^{2}},
\end{equation}
because it takes the particularly simple form
\begin{equation}
\left\{ -\frac{1}{2r}\frac{\partial }{\partial r}r\frac{\partial }{\partial r%
}+V_{r}(r)+\frac{1}{r^{2}}\left[ -\frac{1}{2}\frac{\partial ^{2}}{\partial
\theta ^{2}}+V_{\theta }(\theta )\right] \right\} \psi (r,\theta )=E\psi
(r,\theta ).
\end{equation}
For comparison purposes throughout this paper we follow the notation used by
Alhaidari\cite{A07}.

In order to separate the Schr\"{o}dinger equation into two one-dimensional
eigenvalue equations we write $\psi (r,\theta )=r^{-1/2}R(r)\Theta (\theta )$
where the angular factor is a solution to
\begin{equation}
\left[ -\frac{1}{2}\frac{d^{2}}{d\theta ^{2}}+V_{\theta }(\theta )\right]
\Theta (\theta )=E_{\theta }\Theta (\theta ),  \label{eq:Schr_ang}
\end{equation}
with the boundary condition $\Theta (\theta +2\pi )=\Theta (\theta )$. The
remaining radial equation is
\begin{equation}
\left[ -\frac{1}{2}\frac{d^{2}}{dr^{2}}-\frac{1/4-2E_{\theta }}{2r^{2}}%
+V_{r}\right] R(r)=ER(r).  \label{eq:Schr_rad}
\end{equation}
Note that the term $-1/(8r^{2})$ is missing from the equation (1.4b) of
Alhaidari's paper. As we will see below it has a dramatic effect on the
condition for the existence of a bound state in the system.

When $V_{r}=0$ equation (\ref{eq:Schr_rad}) becomes the eigenvalue
equation for an attractive square potential when $\alpha
=1/4-2E_{\theta }>0$. It is well known that this operator is
self-adjoint when $0\leq \alpha \leq 1/4$ but it does not support
negative eigenvalues for such values of the strength
parameter\cite{GR93}. On the other hand, when $\alpha >1/4$
($E_{\theta }<0$ ) the operator is unbounded from below and
exhibits a ground state with arbitrarily negative energy that is
physically meaningless\cite{GR93}. In-between there is a critical
value $\alpha _{c}=1/4$ that takes place when $E_{\theta }=0$. The
omission of the term $-1/(8r^{2})$ led Alhaidari to the wrong
critical condition $2E_{\theta }=-1/4$.

The charge distribution chosen by Alhaidari\cite{A07} exhibits zero total
charge and zero dipole moment. The first nonvanishing contribution to the
multipole expansion is the quadrupole term that he wrote in dimensionless
form as
\begin{equation}
V_{\theta }(\theta )=-4\xi \sin (2\theta ),  \label{eq:V_theta}
\end{equation}
where $\xi $ is proportional to the strength of the quadrupole. If we take
into account that $\sin (2\theta -\pi /2)=-\cos (2\theta )$ we can rewrite
the angular equation as
\begin{equation}
\Theta ^{\prime \prime }(\theta )+\left[ 2E_{\theta }-8\xi \cos (2\theta
)\right] \Theta (\theta )=0
\end{equation}
that has exactly the form of the Mathieu equation when
$a=2E_{\theta }$, $q=4\xi $\cite{AS72}. The Mathieu equation
exhibits four types of solutions that are relevant to our problem:
even and odd and each one of period $\pi $ and $2\pi $. They can
be obtained as Fourier series with coefficients that satisfy
well-known three-term recurrence relations\cite{AS72}.

Alhaidari\cite{A07} also derived a three-term recurrence relation
(equation (2.8)). However, his basis set of polynomial functions
$T_{n}(x)$ of $x=\sin (2\theta )$ can at most account for the
solutions of period $\pi $. We verified that the eigenvalues given
by that recurrence relation do not agree with those coming from
the recurrence relations for the coefficients of the Fourier
series\cite{AS72}.

In order to obtain the critical values of the quadrupole moment we
just find the values of $q=q_{c}$ such that $a(q_{c})=0$. Since
the lowest eigenvalue $a_{0}(q)$ is negative for all values of $q$
and vanishes at $q=0$ we conclude that there will be solutions to
the Schr\"{o}dinger equation with negative energy for all values
of the quadrupole strength $\xi $. On the
other hand, Alhaidari concluded that the minimum quadrupole parameter is $%
\xi \approx 0.2557$.

Table~\ref{tab:xi_c} shows the first critical values $\xi _{c}=q_{c}/4$ of
the quadrupole moment obtained from the roots of the eigenvalues $%
a_{m}(q_{c})=0$ and $b_{m+1}(q_{c})=0$, $m=0,1,\ldots $, of the
Mathieu equation\cite{AS72}. We observe the occurrence of pairs of
close critical parameters that are due to the fact that
$b_{m+1}-a_{m}$ vanishes exponentially $q\rightarrow \infty
$\cite{AS72}. This behaviour is not found in the case of a charged
particle in the field of a nonrelativistic electric dipole in
three dimensions because the eigenvalues of the corresponding
angular equation do not exhibit such pairing property\cite{AB08}.

\section{Conclusions}

\label{sec:conclusions}

In this paper we revisited the quantum-mechanical problem of a charged
particle in the field of an electric quadrupole in two dimensions and
obtained results that are quite different from those obtained some time ago
by Alhaidari\cite{A07}. In the first place we found that a missing term in
the radial equation has a considerable effect on the condition for the
existence of negative eigenvalues. In the second place we showed that the
solution of the angular part of the Schr\"{o}dinger equation can be
rewritten as a Mathieu equation which enables us to exploit all the
analytical properties of such well-known equation. By means of the standard
three-term recurrence relations for the coefficients of the Fourier
expansions\cite{AS72} we obtained the critical values of the quadrupole
moment. Such recurrence relations may at first sight resemble the one
derived by Alhaidari for the coefficients of the expansion in ``improved
ultra-spherical Gegenbauer polynomials''. However, the latter recurrence
relation fails to yield the well-known eigenvalues of the Mathieu equation
that appear in the standard tables\cite{AS72}.

Our calculations show that there is no critical quadrupole-moment
strength for the capture of the charge. This fact makes present
quadrupole problem different from the charge in the field of an
electric dipole in three dimensions where such critical value
already exists\cite{BBR67}.

Another important difference between the quadrupole and dipole problems is
the occurrence of pairs of close critical quadrupole strengths that do not
appear in the latter three-dimensional problem\cite{AB08}.

\begin{table}[H]
\caption{First critical values $\xi_c$ of the quadrupole moment. }
\label{tab:xi_c}
\begin{center}
\par
\begin{tabular}{lD{.}{.}{12}}
\multicolumn{1}{l}{Eigenvalue} &  \multicolumn{1}{c}{$\xi_c$} \\ \hline
$a_0$ &   0                    \\
$b_1$  &  0.2270115834     \\
$a_1$  &  1.878402574     \\
$b_2$  &  1.894922593     \\
$a_2$  &  5.324657803      \\
$b_3$ &   5.325793406      \\
$a_3$ &   10.48179309      \\
$b_4$ &   10.48186048      \\
$a_4$ &   17.35709457     \\
$b_5$ &  17.35709827     \\

\end{tabular}
\end{center}
\end{table}


\begin{thebibliography}{9}
\bibitem{A07}  Alhaidari A D 2007 \textit{J. Phys. A} \textbf{40} 14843.

\bibitem{CEFGC01}  Camblong H E, Epele L N, Fanchiotti H, and Garc\'{i}a
Canal C A 2001 \textit{Phys. Rev. Lett.} \textbf{87} 220402.

\bibitem{GR93}  Gupta K S and Rajeev S G 1993 \textit{Phys. Rev. D} \textbf{%
48} 5940.

\bibitem{AS72}  Abramowitz M and Stegun I A 1972 \textit{Handbook of
Mathematical Functions} (Dover, New York).

\bibitem{BBR67}  Byers Brown W and Roberts R E 1967 \textit{J. Chem. Phys.}
\textbf{46} 2006.

\bibitem{AB08}  Alhaidari A D and Bahlouli H 2008 \textit{Phys. Rev. Lett.}
\textbf{100} 110401.
\end{thebibliography}
\end{document}